\documentclass[conference]{IEEEtran}
\usepackage{amsmath,amsfonts}

\usepackage{graphicx}
\usepackage{textcomp}
\usepackage{colortbl}
\usepackage{float}
\usepackage{hyperref}
\usepackage{algorithm}
\usepackage{algpseudocode}
\usepackage[table,xcdraw]{xcolor} 
\definecolor{softgreen}{RGB}{198, 239, 206} 
\definecolor{softred}{RGB}{255, 204, 203}   
\usepackage{comment}
\usepackage{booktabs}
\usepackage{tablefootnote}
\usepackage{xcolor}  
\usepackage{pifont}
\usepackage{float}
\usepackage{fvextra}

\usepackage{longtable}
\usepackage{makecell}
\usepackage{url}
\usepackage{subcaption}
\usepackage{tabularx} 
\usepackage{tabularray} 
\usepackage{multirow} 
\usepackage{multicol} 
\usepackage{placeins}
\usepackage{array}
\usepackage{xspace}
\newcommand{\cmark}{\text{\ding{51}}}
\newcommand{\xmark}{\text{\ding{55}}}
\definecolor{verylightgray}{gray}{0.95}
\newcommand{\shortname}{\textsc{InvarLLM}\xspace}

\usepackage{listings}
\usepackage{xcolor}
\usepackage{tcolorbox}
\usepackage{enumitem}

\ifCLASSINFOpdf
\else
\fi
\begin{document}
%
\title{\shortname: LLM-assisted Physical Invariant Extraction for Cyber-Physical Systems Anomaly Detection}

\author{
\IEEEauthorblockN{Danial Abshari}
\IEEEauthorblockA{
University of North Carolina at Charlotte\\
dabshari@charlotte.edu
}
\and
\IEEEauthorblockN{Peiran Shi}
\IEEEauthorblockA{
University of North Carolina at Charlotte\\
pshi@charlotte.edu
}
\and
\IEEEauthorblockN{Chenglong Fu}
\IEEEauthorblockA{
University of North Carolina at Charlotte\\
chenglong.fu@charlotte.edu
}
\and
\IEEEauthorblockN{Meera Sridhar}
\IEEEauthorblockA{
University of North Carolina at Charlotte\\
msridhar@charlotte.edu
}
\and
\IEEEauthorblockN{Xiaojiang Du}
\IEEEauthorblockA{
Stevens Institute of Technology\\
xdu16@stevens.edu
}
}

\IEEEoverridecommandlockouts
\makeatletter\def\@IEEEpubidpullup{6.5\baselineskip}\makeatother
\IEEEpubid{\parbox{\columnwidth}{
		Network and Distributed System Security (NDSS) Symposium 2025\\
		24-28 February 2025, San Diego, CA, USA\\
		ISBN 979-8-9894372-8-3\\
		https://dx.doi.org/10.14722/ndss.2025.[23$|$24]xxxx\\
		www.ndss-symposium.org
}
\hspace{\columnsep}\makebox[\columnwidth]{}}

\maketitle

\begin{abstract}
Cyber-Physical Systems (CPS) support critical infrastructures such as water treatment and power distribution, but their interaction with the physical world makes them vulnerable to cyber-physical attacks. Invariant-based anomaly detection, which identifies consistent physical or logical relationships (invariants) among system variables, has proven effective in detecting violations of these physical laws in CPS data. However, existing methods are limited: data-driven approaches lack semantic context, while physics-based models require extensive manual engineering.
Our key insight is that large language models (LLMs) can extract rich semantic information from CPS deployment documentation, capturing procedural logic and operational constraints not easily discoverable from data alone. Based on this, we propose \shortname, a hybrid framework that integrates semantic extraction and data-driven evaluation to automatically generate accurate physical invariants. \shortname first uses LLMs to interpret textual documentation and extract the  hypothetical physical invariants in fuzzy function-form representation that indicates potential relationships among system variables. These hypotheses are then evaluated against real system logs using a PCMCI+-inspired K-means validation method, which filters out hallucinated or spurious invariants that do not conform to observed data patterns. This combined approach leverages the semantic grounding of LLMs and the empirical rigor of time-series validation, ensuring both interpretability and reliability. We evaluate \shortname on two real-world CPS datasets—SWaT and WADI—achieving 100\% precision in anomaly detection without generating any false alarms. Our results show that \shortname outperforms all existing invariant-based and data-driven methods, achieving state-of-the-art performance in both accuracy and reliability. This demonstrates the power of integrating LLM-derived semantics with statistical validation, offering a scalable and highly dependable solution for CPS security.

\end{abstract}


%
\IEEEpeerreviewmaketitle

\section{Introduction}



Cyber-physical systems (CPS) are the complex integration of computational resources, physical processes, and network operations. This technology significantly improves the efficiency of the operation of critical infrastructure but also makes them vulnerable to various cyber attacks\cite{b21,b10}. To cope with these threats, the \textit{Physics-based Anomaly Detection} has been proposed to profile a CPS's normal behaviors using statistical or machine learning methods and then detect anomalies when the system behavior deviates from the learned profile. In this detection process, researchers aim to profile the systems' normal behavior in the form of \textit{Physical Invariants}\cite{b22}, which refer to the constant system properties and correlations among different sensors and actuators. These physical invariants are constant system properties and correlations between sensors and actuators that are governed by physical laws and environmental factors. 

Due to this constant and consistent nature, physical invariants are ideal as a CPS normal behavior, which can be used as a baseline for anomaly detection. Cyber-attacks such as sensor reading forgery and fake commands will inevitably violate the related physical invariants and then be detected. For instance, SAVIOR~\cite{b22} study the physical invariants among wheel rotation speeds, steering angles, and acceleration, which reflect stable relationships in the AV's dynamics and control. By establishing a baseline model of these invariants, the system can detect anomalies indicative of cyber-physical attacks, like sensor spoofing or control interference.



Although the physical invariants are proved to be very effective in detecting CPS anomalies\cite{b23}, the process of extracting them is a non-trivial task that requires case-by-case analysis by experts with domain-specific knowledge (e.g., the operation of a water treatment system). Since the architectures and applications of the CPS are highly diverse, it is very difficult to find a systematic way to discover and extract physical invariants from heterogeneous CPS deployments. This significantly limits the scalability and usability of physical invariant-based anomaly detection. While some studies, such as Feng et al.~\cite{b29}, use statistical methods to find correlations among device data, they overlook the semantic meaning of devices and physical processes. Without the guidance of domain-specific knowledge, these methods need to binarize numeric data to reduce the search space, resulting in coarse-grained invariants that attackers may circumvent.

In this work, we aim to automate the extraction of physical invariants for CPS anomaly detection. Our insight is that the documentation accompanying CPS deployments typically includes text, tables, and diagrams that illustrate the constituent components and physical interactions among them. This semantic information is highly valuable for assisting in the extraction of physical invariants, yet it has largely been overlooked by existing anomaly detection solutions. Based on this insight, we propose to automate the extraction of physical invariants by leveraging the power of pretrained large language models (LLMs).

Our method proceeds in three stages.
First, we build an LLM-driven extraction pipeline. A customized document parser feeds the processed CPS document into a chain-of-thought prompt template that is tailored to the typical structure and content of CPS documentation, and the model generates hypothetical physical invariants expressed in a fuzzy functional-form template. This step maximizes the semantic value of the documentation while keeping the output generalizable. Next, we develop a lightweight yet effective causal test algorithm to validate these hypothetical invariants using the historical training data and exclude those invariants that are inconsistent with the physical ground truth.
Finally, we apply the fuzzy invariants to power a sliding window-based regression detector. When we have both the benign and attack data consecutively accommodated in a time window, we should be able to see a significant increase in the regression error due to the discrepancy in the correlation pattern before and after the occurrence of an attack, which triggers alarms. The method maintains lightweight effectiveness and efficiently mitigates false alarms arising from concept drift and inaccurate invariants.



To demonstrate the effectiveness of our proposed method, we conducted a comprehensive evaluation on two popular public CPS security datasets. 
Our approach successfully identified and validated some physical invariants, which were used to detect anomalies. The results confirm that our method achieved a high detection rate with 100\% precision, correctly identifying 29 true positives for SWaT and 13 for WADI, with 0 false positives, despite the complexity of the dataset. Additionally, the method demonstrated scalability by integrating invariants derived from generative AI models and robustness in detecting various attack scenarios. These findings highlight the potential of our approach to enable automated, reliable, and cost-effective anomaly detection across diverse CPS environments

Our contributions are summarized as follows:

\begin{itemize}
    \item \textbf{LLM–driven physical invariant extraction.}  We present the first LLM workflow containing a document parser and a tailored chain-of-thought prompt template aligned with common characteristics of CPS documentation, enabling the effective extraction of physical invariants directly from unstructured CPS documents and eliminating the reliance on manually constructed physics models.

    \item \textbf{Fuzzy functional-form template.} We introduce a symbolic invariant representation that is linear in the coefficients yet accommodates nonlinear bases.  The template retains physical meaning while remaining drift-tolerant and easy to fit online.

    \item \textbf{Hybrid invariant validation pipeline.} We design a PCMCI+-inspired causal discovery algorithm that quantifies empirical dependencies for each hypothetical invariant presented in fuzzy form. We also design an unsupervised K-means filter that derives objective thresholds, yielding a scalable screening step that suppresses LLM hallucinations.

    
    

\end{itemize}

The rest of the paper is organized as follows. In Section~\ref{sec:background}, we describe the background of CPS and physical invariants. In Section~\ref{sec:models}, we present the system model and the threat model. We present our method of invariants extraction and anomaly detection in Section~\ref{sec:methods}. In Section~\ref{sec:evaluation} we present the procedure and results of evaluation.  We discuss related work in Section~\ref{sec:related}. We conclude the paper in Section~\ref{sec:conclusion}.

\section{Background}\label{sec:background}

\subsection{CPS and Documentation}
CPS represents a seamless integration of computational algorithms with physical processes, enabling intelligent control and automation in critical sectors such as energy, transportation, and manufacturing. These systems rely on interconnected networks of sensors, actuators, and control units to facilitate real-time data exchange and decision-making. Unlike traditional industrial automation, which operated in isolated environments, CPS leverages advanced communication technologies to enhance efficiency, scalability, and adaptability ~\cite{b35,b38}.

Since CPS deployments are inherently complex, heterogeneous, and safety-critical, comprehensive documentation is indispensable for their development, operation, maintenance, and security assessment. These documents are not merely supplementary materials but essential components required by industrial standards and regulatory frameworks. For example, the U.S. OSHA Process Safety Management regulation (29 CFR 1910.119) explicitly mandates that facilities handling hazardous chemicals maintain comprehensive documentation, including Piping and Instrumentation Diagrams (P\&IDs) and detailed equipment specifications ~\cite{osha2000process}. Similarly, NIST's Guide to ICS Security recommends maintaining complete and up-to-date logical network diagrams of control systems ~\cite{stouffer2014nist}. Industry experts emphasize that thorough documentation is fundamental, noting that ``every single instrument in the plant" should be included in a comprehensive tag list ~\cite{ISAControlDoc}.
 
CPS documentation typically includes several critical components: (1) System architecture diagrams illustrating the interconnections between different components, subsystems, and data flow paths; (2) Detailed component specifications for sensors, actuators, and controllers with their technical parameters and physical locations; (3) Process flow descriptions explaining physical processes such as water treatment stages or chemical dosing procedures; (4) Control logic documentation describing how the system responds to different inputs; and (5) Network topology diagrams showing how devices are interconnected. For example, datasets like Secure Water Treatment (SWaT) and Water Distribution (WADI) include such comprehensive documentation that details all these aspects of their respective systems.

\begin{figure}[t]
    \centering
    \includegraphics[width=1\columnwidth]{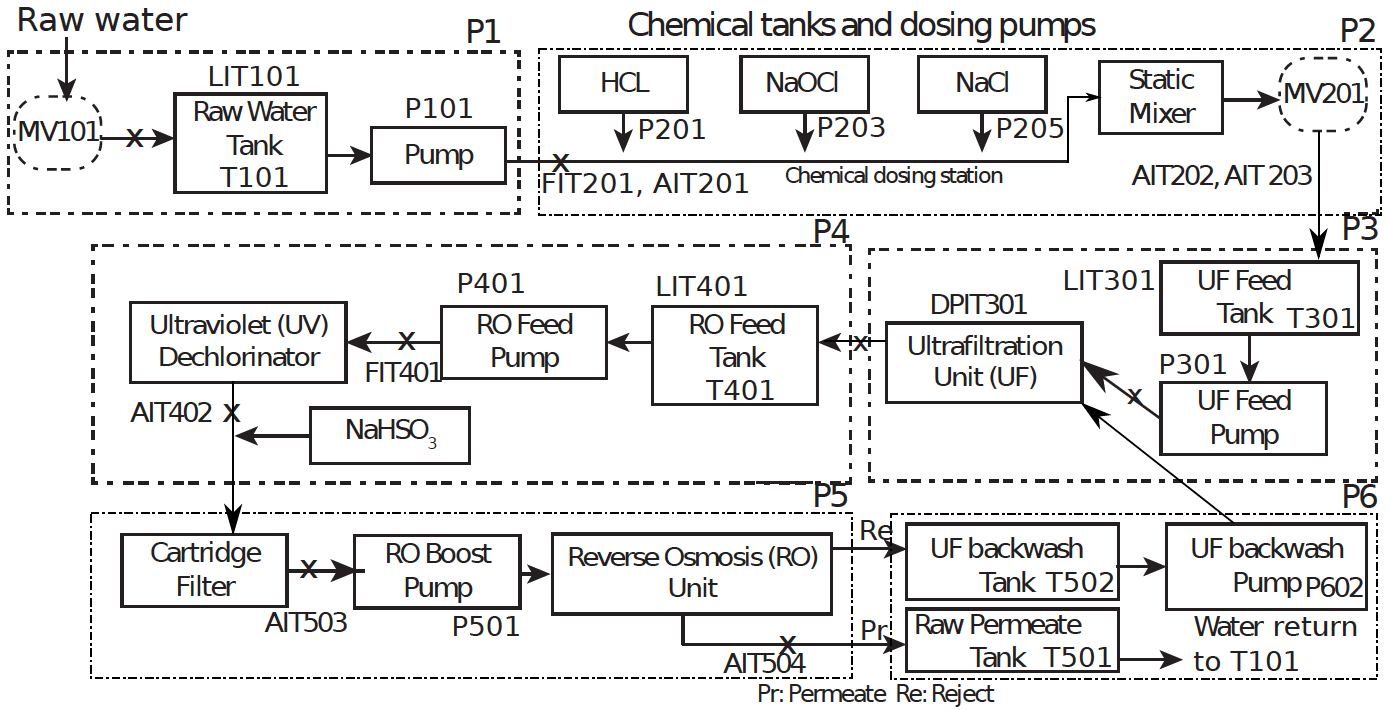}
    \caption{Example of a system architecture diagram from CPS documentation illustrating the physical process flow and component interconnections in the SWaT testbed. The diagram shows water flow paths, sensor placements, and control elements with their semantic relationships \cite{b39}.}
    \label{fig:system-diagram}
\end{figure}

The documentation for systems like SWaT includes detailed P\&ID charts, as shown in Figure~\ref{fig:system-diagram}, displaying water flow paths, chemical injection points, and sensor placements, along with their semantic meanings. For example, a flow sensor labeled $FIT201$ might be documented as "Chemical flow rate at dosing station in Process 2," providing crucial context about its physical significance in the water treatment process. Similarly, level sensors such as $LIT101$ measure water levels in tanks, with documentation explaining their operational ranges and physical interpretation. The diagram illustrates how different processes are interconnected, from raw water storage $(P1)$ through chemical dosing $(P2)$, ultrafiltration $(P3-P4)$, to reverse osmosis $(P5)$, and finally water return $(P6)$.

\subsection{Invariant-based Anomaly Detection}

Invariant-based anomaly detection represents one of the most effective approaches for securing CPS against both cyber attacks and operational faults. This approach leverages physical invariants, mathematical expressions of fundamental physical laws that govern system behavior, to establish a robust baseline for normal operation. These physics-based invariants capture essential relationships between process variables (such as conservation of mass, energy balances, and thermodynamic principles) that must remain consistent regardless of operational conditions. By continuously monitoring these physical invariants, the system can detect subtle deviations that indicate potential anomalies, whether caused by component failures, sensor malfunctions, or malicious attacks. The power of this approach lies in its grounding in immutable physical laws, making it particularly resistant to sophisticated attacks that might evade purely statistical detection methods.

Existing invariant-based approaches can be categorized into several distinct methodologies. Physics-based and manually defined invariants~\cite{paul2013unified,aliabadi2017artinali}, traditionally developed by system engineers during the design phase, benefit from high interpretability as the invariants directly correspond to physical phenomena. However, they face limitations including time-consuming development, inability to capture subtle relationships, extensive domain knowledge requirements, and poor scalability when applied to diverse CPS deployments~\cite{b23}. Data-driven statistical methods~\cite{b29} automatically extract invariants from historical system data, offering systematic identification with less manual effort. Despite these advantages, they suffer from poor interpretability, as the statistically derived invariants frequently lack clear physical meaning and overlook the semantic context of data points. Hybrid approaches~\cite{b11} combine elements of both methods, leveraging machine learning with domain knowledge constraints to balance interpretability with scalability. The effectiveness of invariant-based anomaly detection depends critically on the quality and comprehensiveness of the invariants used. Our work advances the state of the art by automatically extracting physical invariants from system documentation using LLMs, combining semantic understanding with automation benefits while significantly reducing the manual effort required.

\section{System and Threat Model}\label{sec:models}

\subsection{System Model}
In this work, we consider a typical CPS deployment involving physical processes such as water transfer, chemical dispensing, and heat exchange. Control logic resides in controller devices, which issue commands to actuators based on sensor data. All devices periodically report their status to the supervisory control and data acquisition (SCADA) system, where logs are maintained for analysis.

Considering the complexity and scale of typical CPS, the Model-based Design (MBD) paradigm is commonly applied for designing and building CPS systems. In MBD, the design documentation is crucial for the development and maintenance of CPS due to the intricate integration of computational algorithms and physical components that these systems entail. Such documentation ensures that the diverse team of engineers, from software developers to mechanical and electrical engineers, share a common understanding of the system's design goals, functionality, and constraints. 

The content of design documentation for a CPS typically includes several key components. Firstly, a system architecture diagram that visually represents the system's components and their interconnections. Secondly, detailed specifications of each component, including both the physical devices (sensors, actuators, etc.) and the software modules (algorithms, data processing units, etc.). Thirdly, interface descriptions that detail how the components communicate and interact with each other are vital for the integration of heterogeneous elements into a cohesive system. Lastly, operational protocols and user manuals help in understanding how the system is intended to operate under various conditions and how users can interact with it efficiently.

\subsection{Threat Model}

We consider a CPS comprising interconnected sensors, actuators, controllers, and communication networks that rely on real-time data for operational decisions. We assume attackers to be active attackers that have the goal of inducing dangerous physical actions leading to safety hazards, equipment damage, or service disruptions. We assume they have partial or full knowledge of the system's physical and control models, and are capable of compromising one or more components such as sensors, actuators, or communication links. 
The attackers can achieve their goals by manipulating sensor data through false data injection or replay attacks, tampering with actuator commands, hijacking controllers to alter control logic, and exploiting communication channels. We assume attackers cannot compromise all devices in a CPS deployment, as vulnerabilities are typically device-specific—a common assumption in CPS anomaly detection research. Passive attackers who only eavesdrop on network communications to steal information are out of scope for this work.

\section{Design of \shortname}\label{sec:methods}

In this section, we present the design of \shortname, which integrates CPS semantic information and datasets to extract accurate and interpretable physical invariants. We then introduce a novel method that leverages the extracted invariants for anomaly detection. 


\subsection{Overview of \shortname Design Challenges}

\begin{figure*}
\centering
\includegraphics[width=0.95\textwidth]{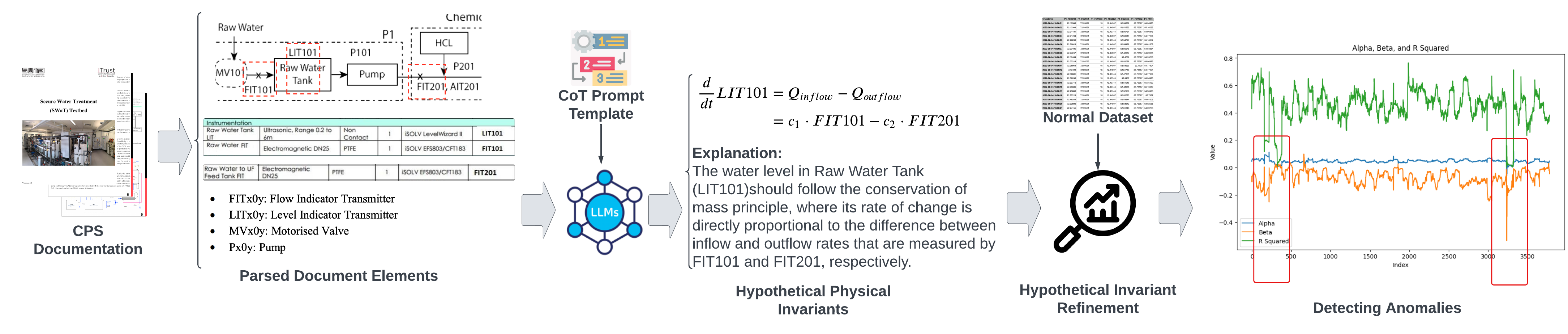}
\caption{The overview of \shortname workflow.}
\label{fig:overview}
\end{figure*}

The workflow of \shortname consists of three primary phases. First, in the \textbf{Hypo-Invariants Extraction} phase, we employ the state-of-the-art large language models (LLMs) to extract hypothetical physical invariants from the semantic content provided by CPS specification documents. Next, the \textbf{Hypo-Invariants Validation} phase refines these hypothetical invariants using statistical analysis on the dataset, without any training. This phase combines semantic information from documentation with statistical evidence from data to filter out invariants that are physically implausible. Finally, in the \textbf{Anomaly Detection} phase, the validated invariants are used to detect anomalies in real-time CPS data streams.

Our design aims to overcome the following challenges: 

\begin{itemize} 

    \item \textbf{Unstructured Documentation.} CPS specification documents typically lack standardized formatting. Unlike source code, these documents are not easily parsed into structured representations, which limits the direct application of techniques such as code analysis used in prior work~\cite{abbas2024sain}.

    \item \textbf{Complexity of Invariants.} CPS operations often involve intricate physical and chemical processes among multiple actuators and sensors. Modeling them precisely is difficult and requires not only expert knowledge but also field investigation and measurements. For instance, estimating water flow in a pipe requires solving fluid dynamics equations based on multiple variables, such as valve position, fluid pressure, and density, which is an inherently complex task.
    
    \item \textbf{Concept Drift.} Although physical invariants are grounded in fundamental laws, real-world CPS environments are subject to varying external factors such as temperature and humidity. These changes introduce concept drift, which can degrade the performance of learning-based methods over time and increase false positives.

    \item \textbf{LLM Hallucination.} LLMs may generate plausible-sounding but incorrect invariants due to hallucination. Moreover, due to their poor performance on structured, low-semantic sequential data~\cite{tan2024language,pang2024uncovering} and limited context window, we cannot input both documentation and datasets simultaneously during invariant generation. This means the LLMs are generating invariants without referring to dataset, which may lead to inconsistencies between inferred invariants and the physical ground truth and thus cause false alarms.
    
    \end{itemize}

To address the first challenge, we design a chain-of-thought prompting strategy that guides the LLM to extract invariants from content commonly found in CPS testbeds and essential for describing the CPS deployment workflow (Section~\ref{subsec:extraction}). This approach improves generalizability across diverse CPS scenarios. For the second and third challenges, we introduce fuzzy representations of physical invariants that capture the \textit{Functional Form} relationships among CPS components without requiring precise mathematical modeling. This abstraction enables the detection of coarse-grained correlations among devices, making the approach more tractable than quantitative modeling and inherently more robust to concept drift (Section~\ref{subsec:extraction} and Section\ref{subsec:validation}). Finally, to mitigate the effects of LLM hallucination, we design a data-driven invariant refinement phase that filters out physically implausible or inconsistent invariants, thereby reducing the likelihood of false alarms (Section~\ref{subsec:detection}).

\subsection{Hypo-Physical Extraction}\label{subsec:extraction}
In this section, we present the workflow for extracting potential physical invariants from CPS documentation using LLMs. The key insight is that LLMs are pretrained on extensive knowledge in physics, engineering, and operational science, enabling them to generalize and infer physical relationships among CPS components based on the deployment descriptions provided in unstructured documentation. To generate invariants suitable for downstream analysis while minimizing hallucinations, it is essential to ensure that the LLM processes the full context of the CPS testbed. The model must also be guided by clear instructions on the reasoning steps and provided with explicit definitions regarding the expected form and type of invariants.

\subsubsection{Documentation Processing} CPS documentation is typically unstructured and lacks a standardized schema. As a result, critical technical details are often scattered across different sections and formats, and may not be clearly explained in the main body text. For instance, in the SWaT testbed documentation~\cite{b39}, the installation locations and control logic of two flow-control valves (FIT101 and FIT201) are presented in system diagrams \ref{fig:overview} and \ref{fig:sample_results}, while additional specifications appear in tables \ref{tab:dataset_details} and \ref{tab:Types_of_LLMs}. If any of this contextual information is omitted, LLMs are unlikely to generate correct physical invariants, even when provided with related textual descriptions. Moreover, CPS documents are often shared as PDF files with inconsistent formatting and disorganized structure. In the case of the SWaT specification, for example, tables are rendered using a mixture of text and embedded images for device labels. This hybrid format hinders the ability to associate device specifications with their corresponding visual representations in the documentation.

To overcome this challenge, we adopt the open-source document parser \texttt{Unstructured.io}~\cite{unstructured2025} to automatically extract and classify CPS documentation into a collection of text chunks, figures, and tables. Leveraging its built-in Optical Character Recognition (OCR) capabilities, the parser can identify the structure of tables, even when presented as mixed text and image content. These elements are then organized into a Markdown file, where cross-references are established among related components. To ensure completeness, we input the Markdown file and associated images into the LLM’s context and first prompt the model to generate detailed descriptions for each figure and table. This preliminary step not only serves as a validation mechanism but also creates a semantically enriched context for the subsequent invariant extraction phase.

\subsubsection{Physical Invariant Formalization}

To ensure that LLMs reliably generate physical invariants in a consistent format suitable for the subsequent anomaly detection phase, a precise definition of the desired invariant representation is necessary. Existing methods often use residual-error-based invariants, represented as $f(\mathbf{x}) - g(\mathbf{x}) = 0$ with fixed numeric coefficients derived from either training datasets or ontology-based approaches. However, in practice, these numeric coefficients tend to drift due to environmental variations, resulting in chronic false positives or significant recalibration overhead. Alternative approaches, such as those used by HAWatcher~\cite{fu2021hawatcher} and Feng et al.~\cite{b29}, utilize pairwise assertion rules to mitigate coefficient drift; nonetheless, they are comparatively coarse-grained and insufficiently expressive for capturing subtle and complex physical invariants. Consequently, we propose a fuzzy \emph{functional-form} representation of physical invariants that maintains the structural integrity or \textit{shape} of underlying physical laws (e.g., mass balance, energy conservation, kinematic relationships) while keeping coefficients undetermined. This design effectively balances statistical precision, robustness against environmental variations, and scalability. A \emph{functional‑form physical invariant} in \shortname is expressed as a linear composition of potentially non‑linear basis terms:

\begin{equation}
\label{eq:ffi}
\sum_{k=1}^{m} C_k,g_k\bigl(\mathbf{s}(t)\bigr) = 0,\qquad  g_k \in \mathcal{G}, ; C_k\in\mathbb{R},
\end{equation}

In the equation \ref {eq:ffi}, $\mathbf{s}(t)$ denotes the vector of sensor/actuator readings, and $g_k(\cdot)$ represents a \textbf{non‑linear} operation of readings such as derivative, multiplication, or exponentiation. The coefficients $C_k$ enter the equation \textbf{linearly}.  The admissible library $\mathcal{G}$ is restricted to: (i) time derivatives, (ii) exponential terms, (iii) pairwise products, and (iv) domain‑specific transforms that respect dimensional homogeneity.  This ''linear‑in‑parameters'' structure is widely used in polynomial regression and sparse identification of nonlinear dynamics (SINDy), enabling convex estimation and statistical guarantees.
Many conservation laws (mass, charge, energy) rearrange naturally into the form of Eq.\eqref{eq:ffi}. Linear‑parameter models admit closed‑form least‑squares estimates and confidence bounds, simplifying the regression procedure used later for anomaly detection.  By keeping coefficients free, Eq.\eqref{eq:ffi} absorbs slow concept drift without altering the symbolic relationship, thereby preventing false alarms due to normal environmental changes.

\subsubsection{Chain of Thought Prompt Template}
Although CPS documentation exhibits diverse styles regarding content layout and organization, most of them follow a hierarchical narrative structure. Typically, these documents begin with an overview of the procedures and subsequently elaborate on detailed processes individually. Additionally, common elements often found in CPS documentation include diagrams illustrating the topological arrangement of involved devices, descriptions of logical controllers, and more importantly, tables detailing datapoints. These tables usually indicate labels, functions, and units for events and commands, and these labels correspond directly to those within the dataset. Leveraging these consistent attributes, we design the \emph{chain-of-thought} (CoT) prompt template, guiding LLMs to systematically inspect all critical document sections in a deterministic sequence and generate the desired functional-form invariants. Based on the above insight, we create the prompt template as detailed in listing below.

\begin{figure}[t]

\begin{tcolorbox}[
    title = \bfseries CoT Prompt Template for Extracting Invariants,
    colback = gray!07,
    colframe = black,
    fonttitle = \sffamily,
    coltitle = white,
    colbacktitle = black
]
\small
\label{lst:prompt}

\textbf{Role Assignment:} \\ You are a \emph{CPS Documentation Analyst}.  
Your task is to read the provided CPS documentation and
\textbf{generate functional‑form physical invariants} that can be used for
anomaly detection.

\bigskip
\textbf{Steps of operation}  
\begin{enumerate}

  \item \textbf{Glossary Construction}  
        Extract a list of glossaries for the name of procedures, devices, and operations from headings, tables, figures, and captions in the form (\verb|Canonical\_ID|, \verb|Aliases|, \verb|Description|> where Alias should be the device or procedure labels as presented in the document.  Use only \texttt{Canonical\_ID}s in later steps; mark unknown Aliases and locations as \texttt{UNKNOWN}.

  \item \textbf{Process Summaries}  
        For each process (e.g.\ ``Raw Water Treatment''), write a
        summary and cite the involved devices.

 \item \textbf{Candidate data points Groups}
        In each process, list the groups of data points that are plausibly
        linked by any fundamental physical relationship governing
      the system (e.g.\ conservation principles, flow dynamics,
      thermodynamic or kinematic couplings). Cite the source page/figure/table IDs for every group.

  \item \textbf{Invariant Generation}  
        For every group of datapoints $\{s_1,\dots,s_p\}$, compose physical invariants of the form  
        \[
          \sum_{k} C_kg_k\!\bigl(s_1,\dots,s_p\bigr) = 0,
          \qquad
          g_k\in\mathcal{G}
        \]
        where  
        $\mathcal{G}=\{\,s_i,\;s_i^{k},\;
        s_i  s_j,\;
        s_i/s_j,\:
        \tfrac{d}{dt}s_i,\;
        \tfrac{d^{2}}{dt^{2}}s_i\,\}$.  
        \begin{itemize}
          \item Leave coefficients \(C_k\) \emph{symbolic}; do not assign numbers.  
          \item Perform a dimensional‑homogeneity check to remove invalid
                equations.  
          \item Output the equation, deduction explanation, and Refs of source to each invariant.
        \end{itemize}

\end{enumerate}

\bigskip
\textbf{Placeholders (fill or replace as appropriate)}
\begin{itemize}
  \item \texttt{[Glossary]}
  \item \texttt{[Process Summaries]}
  \item \texttt{[Candidate Groups]}
  \item \texttt{[List of Invariants]}
\end{itemize}

\bigskip
\textbf{Key Rules}
\begin{enumerate}[label=\arabic*.]
  \item \emph{No numeric coefficients}; keep \(C_k\) symbolic.
  \item Use only IDs from the glossary.
  \item Explain each equation in $\le40$ words.
  \item Reject invariants that fail dimensional checks.
\end{enumerate}

\end{tcolorbox}

\end{figure}

In the prompt template, we first instruct the LLM to compile a \emph{glossary} of every sensor, actuator, and process identifier found in the documentation to prevent omissions and aliasing inconsistencies that would otherwise propagate through later reasoning.  Next, we require the output of each step to be accompanied by explicit page/figure/table citations that serve as an auditable information chain that the model can re‑use as context in subsequent
steps, thereby reducing hallucinations. The reasoning workflow is organized  as a hierarchy:
\textit{process} $\rightarrow$ \textit{groups of candidate datapoints} $\rightarrow$ \textit{functional‑form equations}.  Finally, we impose precise syntactic and semantic constraints on every invariant, specifying (i) the admissible basis functions, (ii) symbolic
coefficient notation, (iii) dimensional‑homogeneity checks, and (iv) a short natural‑language rationale.  Collectively, these template ensures that the LLM’s output is complete, self‑consistent, and ready to be used by the following steps in the pipeline.

\subsection{Hypo-invariant Validation}\label{subsec:validation}

While LLMs are effective at uncovering plausible physical relationships, their outputs may include inaccurate or overly complex expressions since LLMs reason without direct access to the dataset. Therefore, a rigorous data-driven validation step is essential to remove unreasonable hypothetical invariants that contradict observed system behaviors and would otherwise cause false alarms in the detection phase. Here we combine the PCMCI+-inspired causal discovery algorithm \cite{PCMCIplus} with a K-means-based statistical thresholding method. PCMCI+ scores the equation for each hypothetical invariant by measuring the empirical dependency between its left- and right-hand sides, while unsupervised K-means clustering derives an objective cut‑off that separates meaningful relationships from noise. Together, these two components determine whether a hypothesized invariant captures an underlying physical dependency in the dataset and ensure that only robust equations advance to the anomaly detection phase.

\subsubsection{Causal Validation via PCMCI+ with GPDC}

We incorporate PCMCI+, a causal discovery algorithm designed for multivariate time series data. PCMCI+ is particularly effective for handling high-dimensional, autocorrelated, and nonlinear sensor data commonly found in industrial control systems. The PCMCI+ algorithm can potentially provide unoriented links since the algorithm can detect the Markov equivalence class of contemporaneous links \cite{PCMCI_application}.

For our validation task, we specifically employed the Momentary Conditional Independence (MCI) test from PCMCI+, which focuses solely on contemporaneous dependencies without considering time lags. To perform the MCI test, we used the Gaussian Process Dependency Criterion (GPDC) as the conditional independence test. GPDC is particularly suited to our setting for several reasons. First, it effectively models nonlinear, additive dependencies commonly observed in sensor data. Unlike linear correlation measures, GPDC leverages Gaussian processes to flexibly capture a wide range of dependency structures, including heteroscedastic or nonmonotonic relationships, without assuming a specific functional form. Second, GPDC maintains good sensitivity even in small-sample scenarios, where traditional methods often struggle. This ensures that the validation remains robust even in short segments of time series or under system transients.

Although PCMCI+ can theoretically evaluate the correlations for every possible combination of variables, this exhaustive search is impractical for CPS datasets for two main reasons:
(1) The computational complexity of building a full causal graph is prohibitively high for high-dimensional time series datasets (e.g., WADI dataset contains 123 devices, which stand for 123 dimensions). When considering multiple devices, the number of potential conditional independence tests and the size of their corresponding conditioning sets grow combinatorially; 
(2) Applying PCMCI+ across all variables simultaneously can introduce network-wide instability. Due to the dependency-based nature of the algorithm, adding or removing a single variable can alter the conditioning sets and, consequently, affect which causal links are discovered. This sensitivity can lead to inconsistent results when the variable set changes, making the validation of specific invariants less reliable in a full-graph context. This results in both practical infeasibility and a substantial risk of overfitting or statistical power degradation.

Our objective is to validate individual fuzzy function-form physical invariants rather than to reconstruct a full causal graph of the system. We can apply PCMCI+ correlation search in a pairwise fashion, where we aim to assess the strength of correlation relationships among elements indicated in each specific invariant. For a hypothetical invariant, we first construct new time series derived from the invariant equations that involve a combination of device readings. Then, we compute both sides of the equation to generate a pair of synthetic time series. For example, given a hypothetical invariant:

\begin{equation}
\label{equ:example}
\frac{d}{dt} LIT101 = C_1 \cdot FIT101 - C_2 \cdot FIT201
\end{equation}

We estimated the time derivative of \textit{LIT101} and created a new signal by computing the right-hand side (i.e., the weighted sum of \textit{FIT101} and \textit{FIT201}). These two resulting time series, representing both sides of the invariant, were then tested using the MCI framework in PCMCI+ with GPDC as the dependency criterion.

The output of the GPDC-based PCMCI+ test includes both a test statistic score and a p-value. While the p-value is used for assessing statistical significance, the test statistic score itself (in this case, the PCMCI+ score) is more informative for evaluating the strength of the conditional dependence between the left-hand side and right-hand side of the invariant equation. A larger score indicates a stronger contemporaneous dependence between the constructed variables, suggesting that the proposed invariant captures a real relationship between the sensors more accurately. This interpretation is particularly useful in our validation context, where the goal is not only to test for significance, but also to compare and rank the reliability of different invariants based on the strength of their underlying dependencies.

By applying PCMCI+ in a focused, pairwise fashion, we are able to measure the degree of correlation between the left‑ and right‑hand sides of the equation with GPDC kernel. Compared to exhaustive search, our method incorporates statistical rigor and causal insight while avoiding the combinatorial explosion of a full‐graph search.

\subsubsection{Threshold Selection via K-means Clustering}

To statistically determine the threshold of scores for accepting or rejecting hypothetical invariants, we employ K-means clustering on the set of PCMCI+ scores obtained from all validated invariants. Specifically, we applied K-means with $k=2$ to partition the dependency scores into two clusters: one representing strong dependencies and the other weak or spurious dependencies.

The K-means algorithm iteratively minimizes the within-cluster sum of squares, as defined by:
\begin{equation}
\underset{C}{\arg\min} \sum_{i=1}^{k} \sum_{\mathbf{x} \in C_i} \|\mathbf{x} - \boldsymbol{\mu}_i\|^2
\end{equation}

\noindent where $C = \{C_1, C_2, \ldots, C_k\}$ represents the set of clusters, and $\boldsymbol{\mu}_i$ is the mean of points in cluster $C_i$.

For our specific application with $k=2$, we established the threshold value ($\tau$) as the midpoint between the two cluster centroids:
\begin{equation}
\tau = \frac{\boldsymbol{\mu}_1 + \boldsymbol{\mu}_2}{2}
\end{equation}

\noindent where $\boldsymbol{\mu}_1$ and $\boldsymbol{\mu}_2$ are the centroids of the higher and lower value clusters, respectively.

The integration of K-means into the PCMCI+ pipeline offers two key benefits. First, it avoids manually selecting a hard threshold, which may be dataset-specific or arbitrary. Second, it adapts to the natural distribution of dependency scores and allows for robust, unsupervised filtering of weak or non-informative invariant candidates.

\subsection{Anomaly Detection}\label{subsec:detection}

The physical invariants after validation are just a fuzzy approximation of the relationships among involved devices, presented in the function-form format with coefficients undetermined. As a result, they cannot accurately predict some devices' readings from other devices' readings, and we cannot directly apply them for detecting anomalies using the way of residual-error-based methods. Correspondingly, we propose the sliding-window-based regression method for detecting anomalies. Our insight is that, in a linear regression among the components assembled in the function-form linear combination, the regression error should remain constant with no prominent change in the high-level interaction model (e.g., water flow dynamics). However, when attacks intercept or inject values into devices, the correlation will change to a different form. If we have both the benign and attack data consecutively accommodated in a time window, we should be able to see a significant increase in the regression error due to the discrepancy in correlation pattern before and after the occurrence of an attack. 

The validated invariants are intentionally \emph{fuzzy}: they capture only the functional form of the relationship among devices, leaving all coefficients undetermined. Consequently, they cannot be directly used to predict one sensor's reading from others with the precision required for residual‑error detectors. Instead, we introduce a \emph{sliding‑window regression} strategy. Within a time window we fit an ordinary‑least‑squares model to the variables that appear in an invariant. Under normal system operation the regression error remains low and stable because the high‑level physical invariant (e.g., water‑flow dynamics) does not change, but an attack that tampers with sensor or actuator values disrupts this invariant and changes the correlation pattern.
If both benign and tampered data samples fall inside the same window, the regression fit will degrade sharply and produce a marked spike in the sum‑of‑squared residuals (SSR), indicating the occurrence of attacks. Because the invariants contain no fixed coefficients, gradual environmental drift merely shifts the regression parameters without disrupting the underlying correlation structure and therefore does not trigger abrupt changes in error as false alarms, giving the method inherent adaptability and robust drift‑tolerance.

\begin{figure}
    \centering
    \includegraphics[width=0.9\linewidth]{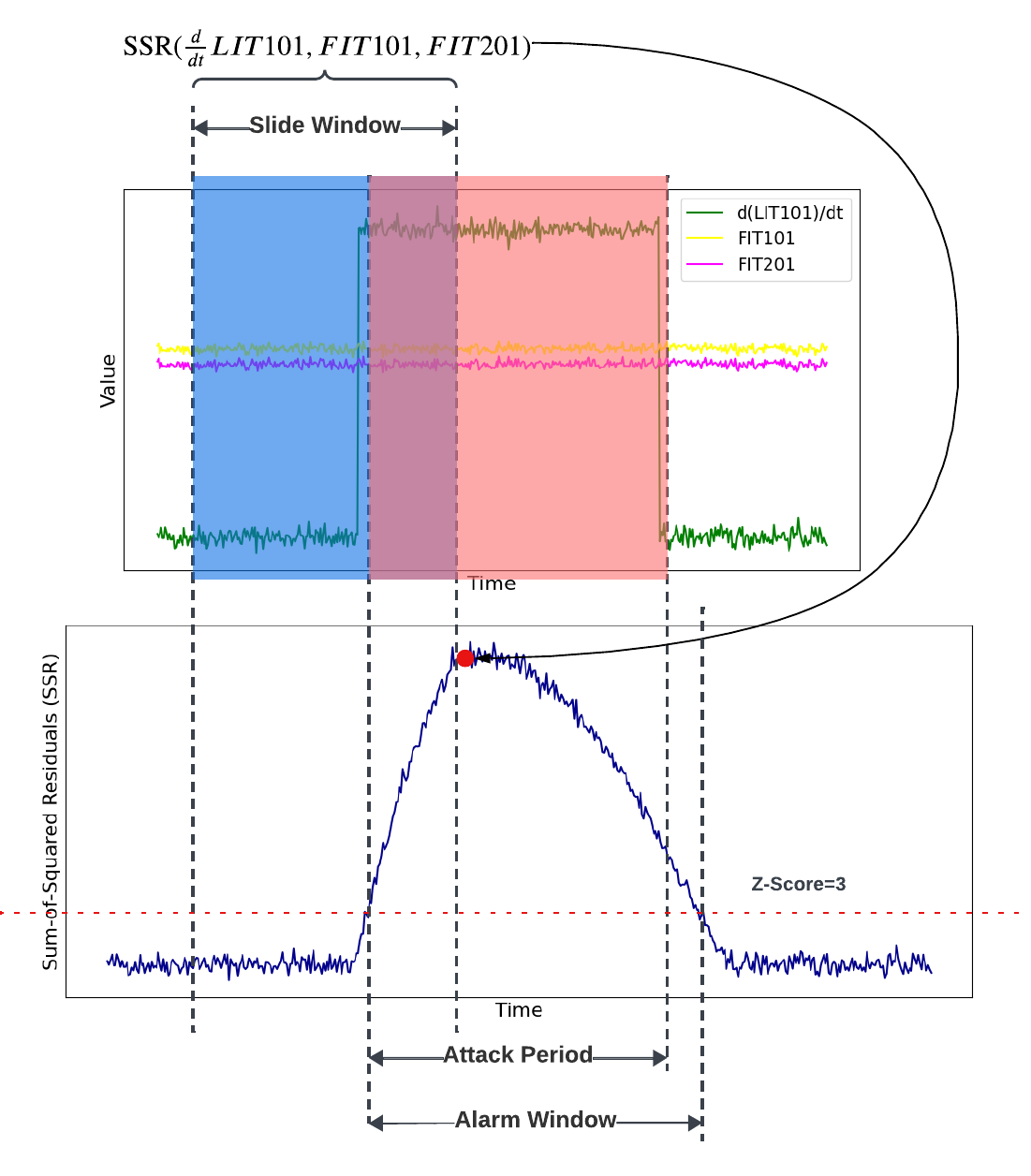}
    \caption{Illustration of detecting anomalous case by calculating the sum of squared residuals (SSR) in sliding windows. The blue line at the bottom contains the list of SSR values of different windows. }
    \label{fig:demo-chart}
\end{figure}

As illustrated in Figure \ref{fig:demo-chart}, we consider the invariant in Equation \ref{equ:example} as an example. If an attacker freezes the flow-rate sensor \textit{FIT101} by replaying stale values, the normal coupling between \textit{FIT101} and the water-level sensor \textit{LIT101} collapses, leaving only the residual relation $\frac{d}{dt} LIT101 = C_2 \cdot FIT201$. In a sliding window (marked by the blue bar) containing clean samples in the first half and tampered samples in the second (marked by the pink bar), a single regression model cannot simultaneously fit both regimes. Consequently, windows that span both benign and tampered samples exhibit a much larger SSR than windows containing only clean data. As the window slides across the timeline and we plot SSR values, an obvious spike appears where the windows overlap the attack period, clearly marking the onset of the intrusion. The detailed detection algorithm is described in the following algorithm. We assign the value of SSR of a sliding window to the ending timestamp of that window. Then, on the plot chart of SSR values along with time, we raise alarms when the Z-score of SSR value goes beyond 3. 



\begin{algorithm}
\caption{\shortname Anomaly Detection}
\begin{algorithmic}[1]
\Require Validated physical invariants $I = \{I_1, I_2, \ldots, I_n\}$, Data stream $D$
\Ensure Set of detected anomalies $A$

\State Initialize anomaly set $A \leftarrow \emptyset$
\For{each time window $W$ in data stream $D$}
    \For{each invariant $I_i \in I$}
        \State Fit linear regression model for $I_i$ within window $W$
        \State Calculate Sum of Squared Residuals (SSR)
        \State Compute Z-score: $Z = \frac{SSR - \mu_{SSR}}{\sigma_{SSR}}$
        \If{$Z > \tau$} \Comment{$\tau$ is threshold, typically 3}
            \State $A \leftarrow A \cup \{(W, I_i)\}$ \Comment{Mark window as anomalous}
        \EndIf
    \EndFor
\EndFor
\State Merge adjacent anomalous windows in $A$
\State \Return $A$
\end{algorithmic}
\end{algorithm}

Sliding window size significantly impacts detection performance and requires careful optimization. A small sliding window will be more sensitive to detect subtle anomalies but could cause more false alarms, while a large window could contain too much clean data that dilutes the anomalous data. To determine the optimal window size for each invariant, we apply it to detect anomalies on the training dataset with different sliding window sizes. Then, we choose the smallest window size that generates no alarms as the optimal window size for that invariant. This method tries to avoid false alarms while maintaining detection sensitivity.

\section{Evaluation}\label{sec:evaluation}

In this section, we evaluate the performance of \shortname for detecting anomalies in CPS deployments.  We begin by describing our experimental setup, followed by the invariant extraction and validation process, and present the detection results.

\subsection{Evaluation Setup}
We evaluate \shortname using two publicly available real-world CPS security datasets: the Secure Water Treatment (SWaT) \cite{b39} dataset and the Water Distribution (WADI) dataset \cite{ahmed2017wadi}. Both datasets are widely recognized benchmarks for cyber-physical security research.

The SWaT dataset represents a fully operational water treatment testbed with six stages, designed to simulate real-world industrial processes. The system consists of 42 sensors and actuators distributed across the six stages, including 24 sensors measuring parameters such as flow, pressure, level, and water quality, along with 18 actuators controlling pumps, valves, and chemical dosing systems. The dataset was collected over an 11-day period: seven days of normal operation followed by four days incorporating 36 different cyber-physical attacks. The SWaT dataset contains 946,722 recorded instances across 51 attributes, with data captured at one sample per second.

The WADI dataset complements SWaT by focusing on water distribution networks. It was collected from a testbed that represents the next stage in the water treatment process after SWaT. The WADI physical system comprises a larger deployment with 123 devices, including 88 sensors and 35 actuators distributed across three main subsystems: primary supply, secondary grid, and return water networks. The dataset spans a 16-day period, with 14 days of normal operation and 2 days containing 15 attack scenarios. The attacks in WADI target different aspects of the distribution network, including pressure manipulation, valve control, and sensor tampering, providing a more complex network of interactions and dependencies.
\begin{table}[h]
\centering

\begin{tabular}{lcccc}
\toprule
\textbf{Dataset} & \textbf{Dimensions} & \textbf{Normal} & \textbf{Attack} & \textbf{Anomaly Ratio} \\
\midrule
SWaT & 51 & 496800 & 449919 & 12.14\% \\
WADI & 123 & 1048571 & 172801 &  5.77\%\\
\bottomrule
\end{tabular}
\caption{Two benchmarked datasets.}
\label{tab:dataset_details}
\end{table}
Together, these datasets allow us to evaluate \shortname across diverse cyber-physical environments with varying complexities and instrumentation densities. For clarity of presentation, we focus our detailed analysis on the SWaT dataset throughout this section and present the combined results for both datasets at the end.

\subsection{Invariant Extraction from LLMs}
We leveraged multiple state-of-the-art LLMs to extract physical invariants from the SWaT and WADI system documentation and then compared their effectiveness. Each LLM received the same set of prompts designed to elicit an understanding of the physical relationships within these CPS systems. We asked about component interactions, process flows, and potential mathematical relationships between sensors and actuators.
As shown in Table~\ref{tab:Types_of_LLMs}, these extracted invariants captured a diverse range of physical relationships, from simple flow balance equations to more complex interactions between chemical dosing and water quality parameters. The subsequent validation process determined which of these hypothetical invariants accurately represent the physical behavior of the system.
\begin{table}[t]
\centering
\begin{tabular}{lcc}
\toprule
\textbf{LLM} & \textbf{SWaT Invariants} & \textbf{WADI Invariants}\\ \hline
GPT-o1 & 40 & 42 \\
Claude 3.7 Sonnet & 16 & 33 \\
GPT-4o & 23 & 20 \\
DeepSeek & 18 & 17 \\
GPT-4.5 & 18 & 19 \\
\hline
\textbf{Total unique} & 80 & 74 \\
\bottomrule
\end{tabular}
\caption{Number of hypothetical invariants extracted from different LLMs}
\label{tab:Types_of_LLMs}
\end{table}
When comparing LLMs for generating invariants, clear differences stood out. GPT-o1 ~\cite{chatgpt2025} and Claude 3.7 ~\cite{claude2025} consistently produced more accurate and meaningful results, showing a strong grasp of physical relationships and system dynamics. GPT-o1 nailed complex flow balances, while Claude 3.7 delivered clean, well-structured invariants. On the flip side, DeepSeek ~\cite{deepseek2025} and GPT-4.5 sometimes included vague variables or messed up dimensional checks, needing extra cleanup. GPT-4o did well, safely, but occasionally missed subtle patterns. Overall, GPT-o1 and Claude 3.7 led the pack with smarter, more reliable outputs.

\subsection{Invariant Validation}

For the hypothetical physical invariants, we further apply the invariant validation method as described in Section~\ref{subsec:validation} to refine them into a collection of 38 and 12 invariants from SWaT and WADI testbeds, respectively.

\subsubsection{PCMCI+-inspired invariant validation}

This approach assesses causal relationships while controlling for confounding factors in time-series data. Each invariant received a score between 0 and 1, with higher values indicating stronger statistical support for the hypothesized physical relationship, as detailed in Table \ref{tab:comprehensive}. This comprehensive table presents the workflow of \shortname, showing the PCMCI+ validation scores for each extracted invariant.

To evaluate the effectiveness of our PCMCI+-inspired validation method, we conducted a series of experiments on hypothetical invariants extracted from the SWaT dataset. Our goal is to assess whether the proposed sensor relationships are statistically supported by the data, using the GPDC-based MCI test to quantify the strength of contemporaneous dependencies. We focus on three key experimental outcomes: 1) Comparing complex invariants with their sub-invariants; (2) Detecting redundant variables; 3) Analyzing computational efficiency in pairwise versus full-graph validation.

\noindent \textbf{Refinement via Sub-Invariant Comparison}
A key advantage of our validation framework is its ability to refine and rank complex invariants by comparing their PCMCI+ scores with those of their constituent sub-invariants. For example, consider a hypothetical invariant $\mathbb{INV}14$ in table \label{tab:invariant_validation}, we also compute PCMCI+ values for the sub-invariants $\mathbb{INV}15$, $\mathbb{INV}16$, and $\mathbb{INV}17$. This allows us to determine whether the inclusion of additional variables genuinely improves the strength of the proposed dependency. In some cases, we find that the complex invariant yields a stronger dependency than any of its sub-components. For instance, the PCMCI+ test reveals a dependency value of 0.541 for $\mathbb{INV}14$, compared to 0.348 for $\mathbb{INV}15$ and 0.453 for $\mathbb{INV}17$. This suggests that the full invariant captures a meaningful interaction that neither sub-invariant captures on its own. Conversely, in other cases, sub-invariants exhibit higher dependency values than the original complex invariant. For example, for the invariant $\mathbb{INV}2$, we observe a score of 0.786 for the combined expression, while $\mathbb{INV}3$ alone yields a stronger score of 0.966. This indicates that $\textit{FIT502}$ may not be a strong contributor and suggests the original invariant may be over-parameterized. These findings demonstrate that PCMCI+-inspired validation can guide invariant simplification, helping to discard unnecessary terms while highlighting the core contributors to each dependency.

\noindent \textbf{Redundancy Elimination}
Beyond ranking invariant strength, our method also supports redundancy detection by identifying variables that contribute no observable dependency. For example, we evaluated the invariant $\mathbb{INV}14$. The overall dependency score between \textit{$\frac{d}{dt}LIT101$} and the full expression is 0.541, and further inspection shows that the score for \textit{$\frac{d}{dt}LIT101$} and \textit{P101} alone is also 0.453, while \textit{P102} shows no detectable link with \textit{$\frac{d}{dt}LIT101$}. This clearly indicates that \textit{P102} is redundant in the context of this invariant. By revealing such silent or non-influential components, the PCMCI+-based validation process helps eliminate noise from model hypotheses and improve interpretability.

\noindent \textbf{Computational Complexity Analysis}
To evaluate the efficiency of our pairwise validation framework, we compare its computational complexity to that of constructing a full causal graph over all sensors using PCMCI+. In the \emph{graph-based approach}, assume there are $N$ variables. Constructing a full causal graph requires conditional independence testing for all variable pairs, and the total number of tests scales as $\mathcal{O}(N^2 \cdot K)$, where $K$ denotes the average number of conditioning set combinations tested per pair. This quadratic growth in complexity with respect to $N$ makes full graph construction computationally expensive and often infeasible for high-dimensional sensor systems like SWaT.

In contrast, our \emph{pairwise validation approach} evaluates each hypothetical invariant independently. For an invariant involving $d$ variables, the number of tests required scales as $\mathcal{O}(d \cdot K)$, where typically $d \ll N$ since most invariants describe relationships between only 2 variables. As a result, the total computational complexity grows linearly with the number of invariants, enabling scalable and efficient validation across large sensor networks.

This comparison highlights that our pairwise PCMCI+ method is not only computationally tractable but also well-aligned with the goal of verifying specific sensor invariants. By focusing on local, contemporaneous dependencies instead of reconstructing a full causal structure, we achieve significant computational savings while maintaining precision in validating physical relationships.

\subsubsection{k-means clustering}

To establish an objective threshold for invariant selection, we applied k-means clustering with k=2 to the PCMCI+ scores. This approach naturally divided the invariants into two clusters, representing significant and non-significant relationships.

\begin{table}[h]
    \centering
    \small
    \setlength{\tabcolsep}{3pt}
    \begin{tabular}{p{1cm}|p{1.4cm}|p{1.4cm}|p{1.4cm}|p{2.6cm}}
        \toprule
        \textbf{Dataset} & \textbf{Higher Centroid} & \textbf{Lower Centroid} & \textbf{Threshold} & \textbf{Cluster Distribution} \\
        \hline
        SWAT & 0.578 & 0.173 & 0.376 & 50.7\%, 49.3\% \\
        \hline
        WADI & 0.768 & 0.209 & 0.488 & 20.0\%, 80.0\% \\
        \bottomrule
    \end{tabular}
    \caption{K-means clustering results for feature importance thresholding}
    \label{tab:kmeans_thresholds}
\end{table}

\begin{table*}
\renewcommand{\arraystretch}{1.2}
\centering
\resizebox{0.9\textwidth}{!}{
\begin{tabular}{c| |c |p{8cm}|c| c|c| l}
\toprule
\rowcolor[HTML]{FFFFFF} 
\textbf{Dataset} &
\textbf{ID} &
\textbf{Invariants} &
\textbf{Value} &
\textbf{Status} &
\textbf{Window Size} &
\textbf{Attacks Detected} \\ \hline

\multirow{17}{*}{\rotatebox{90}{SWaT}} 
& $\mathbb{INV}1$ & $ \text{AIT202} \cdot \text{FIT201} = c_1 \cdot (\text{P203} + \text{P204})$ & No Links! & \xmark & & \\ \cline{2-7}
& $\mathbb{INV}2$ & $ \text{FIT401} = c_2 \cdot \text{FIT501} + c_3 \cdot \text{FIT502}$ & 0.786 & \cmark & 10 & 21, 34, 35 \\ \cline{2-7}
& $\mathbb{INV}3$ & $ \text{FIT401} = c_4 \cdot \text{FIT501}$ & 0.966 & \cmark & 20 & 8, 22, 32, 34, 35 \\ \cline{2-7}
& $\mathbb{INV}4$ & $ \text{PIT501} = c_5 \cdot \text{FIT503} + c_6 \cdot \text{FIT504}$ & 0.823 & \cmark & 10 & 8, 17, 22, 32, 33, 35 \\ \cline{2-7}
& $\mathbb{INV}5$ & $ \frac{d}{dt}\text{LIT401} = c_{7} \cdot \text{FIT401} - c_{8} \cdot (\text{P401} + \text{P402})$ & 0.717 & \cmark & 10 & 13, 20, 22 \\ \cline{2-7}
& $\mathbb{INV}6$ & $ \text{AIT402} = c_{9} \cdot \text{FIT401}$ & 0.602 & \cmark & 50 & 8, 16, 21, 33, 34, 35 \\ \cline{2-7}
& $\mathbb{INV}7$ & $ \frac{d}{dt}\text{LIT101} = c_{10} \cdot \text{FIT101} + c_{11} \cdot \text{FIT201}$ & 0.56 & \cmark & 40 & 3, 4, 5, 6, 15, 23, 28, 31 \\ \cline{2-7}
& $\mathbb{INV}8$ & $ \frac{d}{dt}\text{LIT301} = c_{12} \cdot \text{FIT201} - c_{13} \cdot \text{FIT301}$ & 0.472 & \cmark & 10 & 8, 17, 24, 36 \\ \cline{2-7}
& $\mathbb{INV}9$ & $ \text{FIT501} = \text{FIT502} + \text{FIT503} + \text{FIT504}$ & 0.448 & \cmark & 20 & 8, 17, 22, 33, 35 \\ \cline{2-7}
& $\mathbb{INV}10$ & $ \frac{d}{dt}\text{LIT101} = c_{14} \cdot \text{FIT201}$ & 0.348 & \xmark & & \\ \cline{2-7}
& $\mathbb{INV}11$ & $ \text{FIT502} = c_{15} \cdot (\text{PIT501} - \text{PIT502})$ & 0.21 & \xmark & & \\ \cline{2-7}
& $\mathbb{INV}12$ & $ c_{16} \cdot \text{AIT201} + c_{17} \cdot \text{AIT202} = c_{18} \cdot \text{MV201} + c_{19} \cdot \text{FIT201}$ & 0.186 & \xmark & & \\ \cline{2-7}
& $\mathbb{INV}13$ & $ 0 = \text{FIT601} - c_{20} \cdot \text{P602}$ & 0.123 & \xmark & & \\ \cline{2-7}
& $\mathbb{INV}14$ & $ \frac{d}{dt}\text{LIT101} = \text{FIT101} - c_{21} \cdot (\text{P101} + \text{P102})$ & 0.541 & \cmark & 40 & 8, 16, 21, 32, 33, 34, 35 \\ \cline{2-7}
& $\mathbb{INV}15$ & $ \frac{d}{dt}\text{LIT101} = \text{FIT101}$ & 0.348 & \xmark & & \\ \cline{2-7}
& $\mathbb{INV}16$ & $ \frac{d}{dt}\text{LIT101} = c_{22} \cdot \text{P102}$ & No Links! & \xmark & & \\ \cline{2-7}
& $\mathbb{INV}17$ & $ \frac{d}{dt}\text{LIT101} = c_{23} \cdot \text{P101}$ & 0.453 & \cmark & 10 & 3, 4, 16, 24, 33, 34, 35 \\ \hline

\multirow{5}{*}{\rotatebox{90}{WADI}} 
& $\mathbb{INV}18$ & $ \text{2\_DPIT\_001\_PV} = c_{24} \cdot \text{2\_PIT\_001\_PV} + c_{25} \cdot \text{2\_PIT\_002\_PV}$ & 0.951 & \cmark& 30 & 1, 4, 10, 11, 12, 13\\ \cline{2-7}
& $\mathbb{INV}19$ & $ \frac{d}{dt}\text{2\_FQ\_401\_PV} = c_{26} \cdot \text{2\_FIC\_401\_PV}$ & 0.507 & \cmark & 45& 6,14\\ \cline{2-7}
& $\mathbb{INV}20$ & $ \text{2\_PIT\_003\_PV} = c_{27} \cdot \text{2\_FIT\_003\_PV}$ & 0.572 & \cmark &35 & 4, 5, 10\\ \cline{2-7}
& $\mathbb{INV}21$ & $ \text{2\_PIT\_002\_PV} = c_{28} \cdot \text{2\_FIT\_002\_PV}$ & 0.455 & \xmark & & \\ \cline{2-7}
& $\mathbb{INV}22$ & $ \frac{d}{dt}\text{1\_LT\_001\_PV} = c_{29} \cdot \text{1\_FIT\_001\_PV} + c_{30} \cdot \text{3\_FIT\_001\_PV}$ & 0.212 & \xmark& & \\ \bottomrule

\end{tabular}
}
\caption{A portion of the physical invariants extracted from SWaT and WADI documents. We also list their validation scores and the capabilities of detecting certain attacks.}
\label{tab:comprehensive}
\end{table*}

The K-means clustering analysis on the SWaT dataset revealed a notably balanced split of invariants between the high- and low-significance clusters (50.7\% vs 49.3\% ). This distribution suggests that the physical relationships in water treatment systems span a more continuous spectrum of statistical significance. The resulting threshold of 0.386 for SWaT provided an objective criterion for invariant selection, allowing us to focus on the most statistically reliable relationships for anomaly detection. Based on these thresholds, we accepted $\mathbb{INV}1$ through $\mathbb{INV}5$ for the SWaT dataset, which showed strong statistical relationships consistent with the underlying physical processes. Conversely, $\mathbb{INV}6$ through $\mathbb{INV}9$ were rejected due to weak causal relationships or inconsistent behavior during validation.

K-means clustering also enabled us to identify borderline invariants with potential for refinement. For example, an invariant similar to $\mathbb{INV}6$ was reformulated by incorporating flow balances from other connected pipes, which improved its PCMCI+ score from 0.348 to 0.712, elevating it above the threshold and making it suitable for inclusion in the final set.

The validated invariants represent diverse physical relationships within the water treatment system:
\begin{itemize} 
\item Flow balance relationships ($\mathbb{INV}1$, $\mathbb{INV}5$, $\mathbb{INV}18$): These express the conservation of flow across components, where inflows and outflows must balance either instantaneously or through a linear relationship. \item Pressure-flow correlations ($\mathbb{INV}2$, $\mathbb{INV}20$): These capture how pressure measurements relate to flow rates, often through a proportional relationship dictated by pipe characteristics. \item Tank level dynamics ($\mathbb{INV}3$, $\mathbb{INV}4$, $\mathbb{INV}19$): These describe how flow rates influence changes in tank levels, linking integrative (or differential) relationships between sensors. \end{itemize}

Due to the page limit, we present the complete list of validated physical invariants on two testbeds in the Appendix. 


\subsection{Detection Methodology}

We implemented our sliding window regression approach using the validated invariants. For each invariant, we apply the method as described in Section~\ref{subsec:detection} to flag the data samples as being anomalous.


To determine the optimal window size for each invariant, we tested durations from 10 to 65 minutes, selecting the smallest window size that eliminated false positives while maintaining detection sensitivity. This optimization was crucial because different physical processes operate at different time scales, requiring customized window sizes for effective anomaly detection. After computing the Z-scores, our detection process follows a systematic algorithm. For each sliding window, we evaluate every validated invariant by fitting a linear regression model and calculating the Sum of Squared Residuals (SSR). These SSR values are normalized using Z-score transformation: $Z = \frac{SSR - \mu_{SSR}}{\sigma_{SSR}}$, where $\mu_{SSR}$ is the mean SSR across all windows and $\sigma_{SSR}$ is the standard deviation. Windows with Z-scores exceeding our threshold $\tau$ (typically set to 3) are marked as potential anomalies. This threshold was chosen to balance sensitivity and specificity, representing a significant statistical deviation that is unlikely during normal operation. Our analysis revealed optimal window sizes for each validated invariant as shown in Table \ref{tab:comprehensive}, which provides a comprehensive view of the entire invariant processing pipeline from extraction to detection.

\subsection{Detection Results}

We evaluate detection performance on two complementary scales.

\begin{itemize}
    \item \textbf{Sample-level accuracy} treats every timestamped record as an independent instance and asks whether it is labeled consistently with the ground truth in the dataset.  While this metric quantifies raw discriminatory power, it is heavily skewed by class imbalance. Considering that the duration of different attack cases could vary significantly, this evaluation method may give a deceptively high true positive rate if a detector mainly detects anomalous samples in long-duration attack cases but actually fails in other short-term cases. Likewise, a handful of isolated false alarms scattered across millions of benign samples could produce a seemingly low false-positive rate. But each of them will demand operators' attention and response.

    \item \textbf{Case-level accuracy} asks the operator-relevant question: does the detector raise an alarm for each attack case (e.g., the 36 documented SWaT dataset), irrespective of how many individual samples were flagged once the alarm was triggered?  This criterion aligns better with real-world practice, where the goal is to detect an \emph{ongoing incident} for triggering a prompt response from the system operators rather than labeling every single anomalous sensor reading. A brief contiguous burst of correct alarms at the start of an attack case meets this need, and any immediately adjacent false-positive tail is usually tolerated as part of the same event. By collapsing temporally correlated samples into single yes/no outcomes, case-level evaluation prevents ``pin-prick'' detectors with scattered alerts from appearing overly accurate and provides a more realistic measure of operational effectiveness. 
\end{itemize}

Accordingly, we regard case-level metrics as a more realistic and operationally meaningful measure of accuracy, and we report both metrics for completeness.

\subsubsection{Sample-level Accuracy}

\begin{table}[htbp]
    \centering
    \begin{tabular}{c|l|c|c|c|c}
        \toprule
        \textbf{Dataset} & \textbf{Metrics} & \textbf{Precision} & \textbf{Recall} & \textbf{FPR} & \textbf{F1} \\
        \hline
        \multirow{2}{*}{\rotatebox[origin=c]{0}{{SWaT}}} 
        & Raw & 42.54\% & 21.96\% & 4.10\% & 28.97\% \\
        \cline{2-6}
        & Corrected & 98.77\% & 76.36\% & 0.13\% & 86.09\% \\
        \hline
        \multirow{2}{*}{\rotatebox[origin=c]{0}{{WADI}}} 
        & Raw & 42.49\% & 67.83\% & 5.6\% & 52.25\% \\
        \cline{2-6}
        & Corrected & 77.93\% & 67.83\% & 1.18\% & 72.48\% \\
        \bottomrule
    \end{tabular}
    \caption{Sample-level evaluation results.}
    \label{tab:sample-level}
\end{table}

We got the raw detection results for SWaT and WADI datasets at rows 1 and 3 in Table \ref{tab:sample-level}, respectively. Taking SWaT as an example, we detected 28,192 data samples as anomalous in 22 consecutive time segments. Among them, 11,993 are true positives and 16,199 are false positives. As listed in Table~\ref{tab:sample-level}, this results in a precision of 42.54\% and a recall of 21.96\%. The overall false positive rate (FPR) is 4.10\%, and F1 score is 28.97\%.

\begin{figure}[htbp]
  \centering
  \subfloat[]{\includegraphics[width=0.45\textwidth, height=1.6cm,keepaspectratio=false]{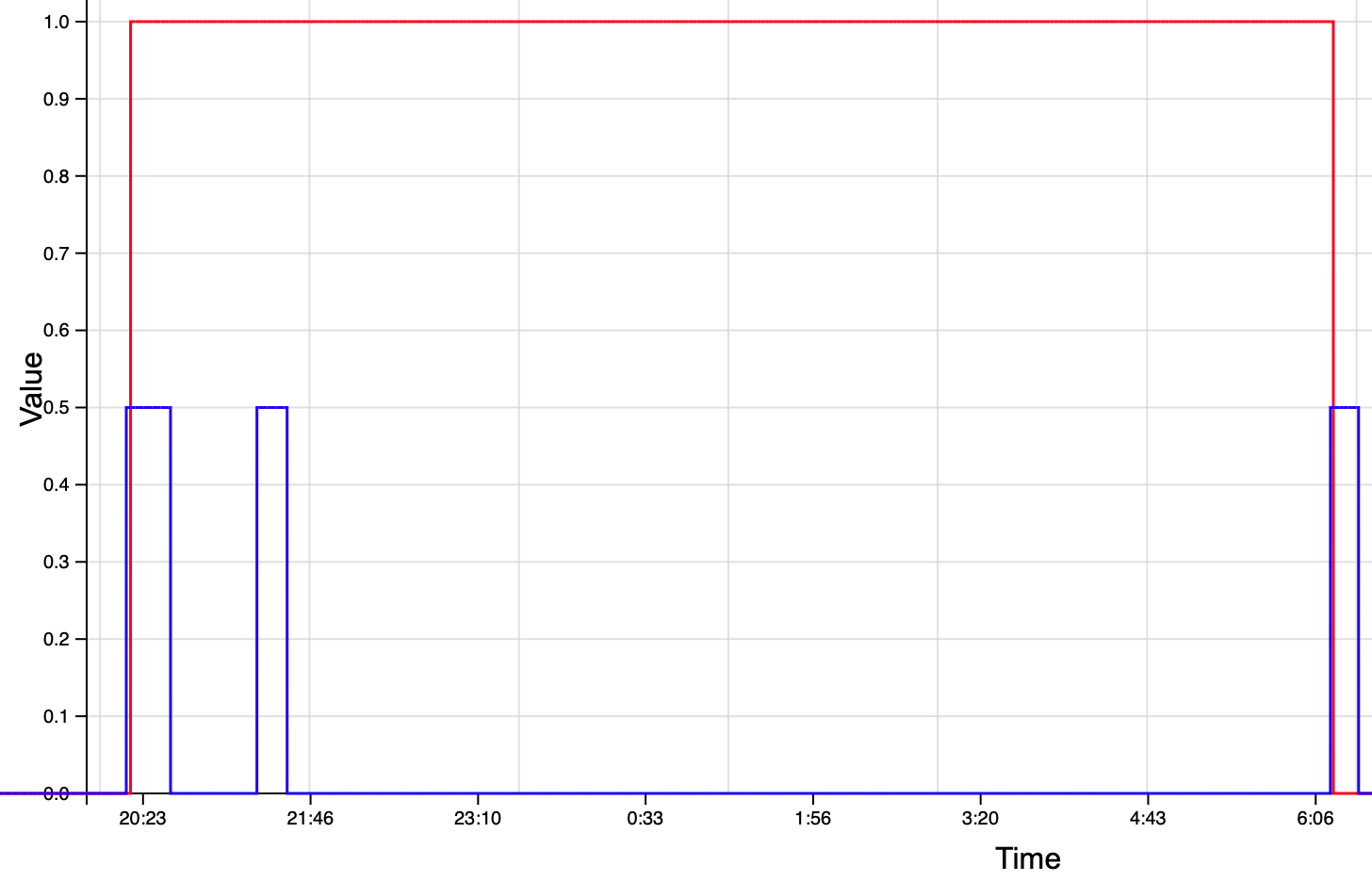}}\\[1em]
  
  \subfloat[]{\includegraphics[width=0.14\textwidth, height=1.6cm,keepaspectratio=false]{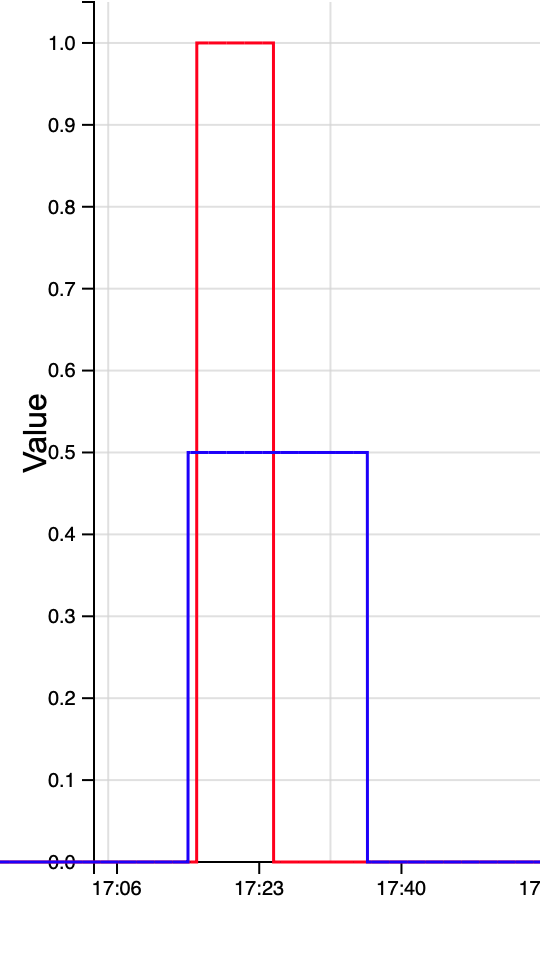}}\hfill
  \subfloat[]{\includegraphics[width=0.14\textwidth, height=1.6cm,keepaspectratio=false]{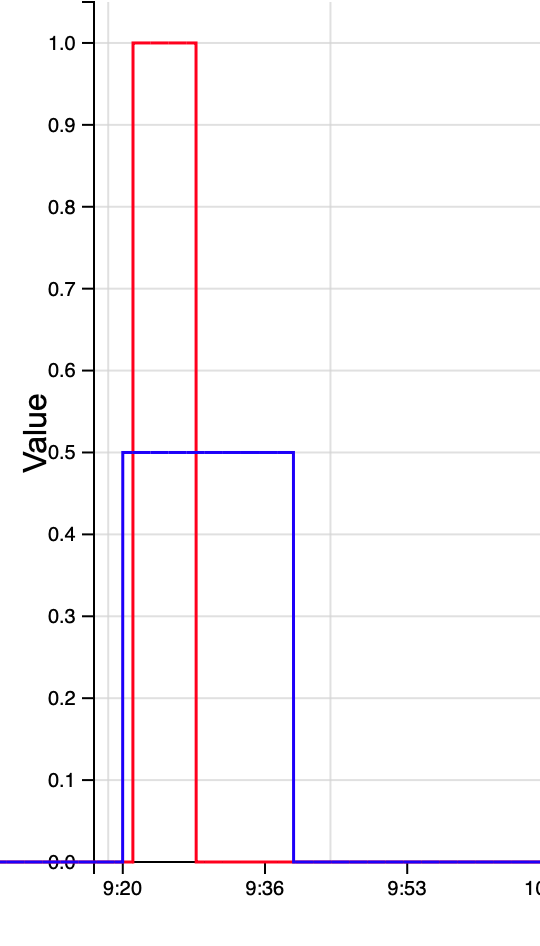}}\hfill
  \subfloat[]{\includegraphics[width=0.14\textwidth, height=1.6cm,keepaspectratio=false]{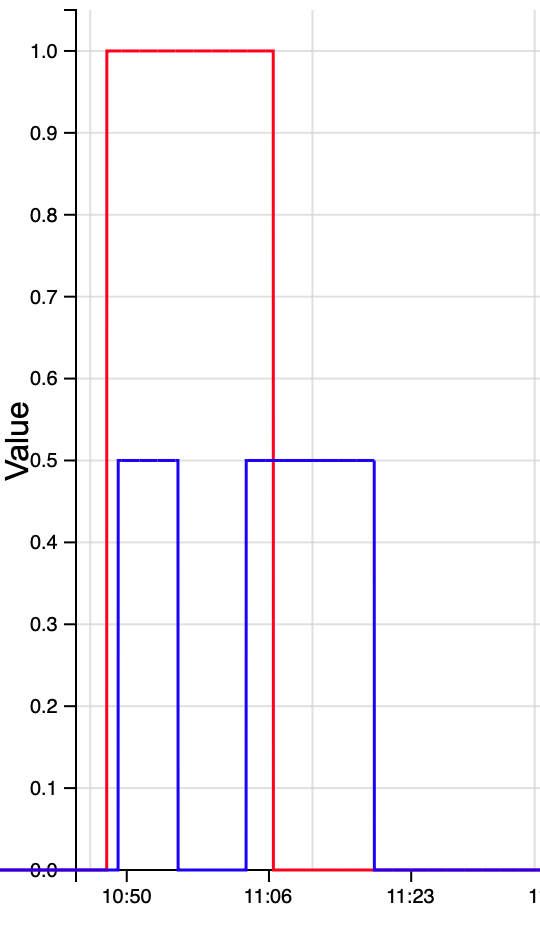}}
  
  \caption{Some examples of detection results on SWaT dataset. The red rectangles stand for periods of attacks, and blue boxes stand for periods of alarms.}
  \label{fig:sample_results}
\end{figure}


Although the raw sample-level results cannot outperform existing solutions on the same dataset, a closer examination of the error distribution reveals three encouraging observations.  
\textbf{First}, \shortname never produces isolated ''pin-prick'' alarms and all 22 alarm bursts are contiguous, ranging from 6 min to 54 min
(average 21.3 min), and every burst overlaps at least one documented attack case.  
\textbf{Second}, there is one unusually long attack that spans longer than 10 hours (i.e., 34,208 anomalous samples, or 62.6\% of all anomalous samples) that explains most of the false negatives. As illustrated in Figure~\ref{fig:sample_results} (b) and (c), once the sliding window sits wholly inside this attack, the mix of benign and malicious samples disappears. So it is reasonable to see the regression error drop, and alarms stop.  Yet \shortname still raises clear alerts at the \emph{start} and \emph{end} of the attack, which precisely causes operators' attention to handle the case regardless of the false negatives in the middle. 
\textbf{Third}, the majority of false positives cluster immediately after true-positive bursts.  Because each window is time-stamped by its \emph{ending} time, early post-attack windows still contain anomalous samples, which inflate the SSR and extend the alarm tail.  In practice, operators can easily recognize this contiguous tail as part of the preceding event, so it imposes negligible nuisance cost. 

Based on these observations, we re-evaluate the performance of \shortname with two corrections on the metrics: 1) we mark all samples of that long attack case as true positives; 2) any alarm raised within one sliding window after an attack ends is not counted as a false positive. By applying these two corrections, we can see the significant improvement in the detection that is presented in Table~\ref{tab:sample-level} with the metrics corrected.

\subsubsection{Case-level accuracy}

As illustrated above, we also evaluate the performance of \shortname using the more reasonable case-level metrics. Instead of checking each individual data sample, we apply the following criteria: 

\begin{itemize}
    \item \textbf{True Positives and False Negatives}: For each attack case, if its period of occurrence overlaps with at least one alarming segment, it is counted as a true positive case. Otherwise it is a false negative case.
    \item \textbf{False Positives}: For each alarming segment, it is counted as a false positive case if it has no overlap with any one of the attack cases. 
\end{itemize}

Based on these criteria, we are able to calculate the precision, recall, and F1-score among the 36 attack cases and 22 alarming segments (SWaT dataset). which gets the result as shown in Table~\ref{tab:performance}

\begin{table}[htbp]
\centering

\begin{tabular}{l|p{1cm}|p{1cm}|c}
\toprule
\textbf{Metric} & \textbf{WADI Dataset} & \textbf{SWaT Dataset} & \textbf{Combined} \\
\hline
Total Attacks & 15 & 36 & 51 \\
True Positives (TP) & 13 & 29 & 42 \\
False Positives (FP) & 0 & 0 & 0 \\
False Negatives (FN) & 2 & 7 & 9 \\
\hline
Precision & 100\% & 100\% & 100\% \\
Recall/Detection Rate & 86.67\% & 80.56\% & 82.35\% \\
F1 Score & 92.86\% & 89.25\% & 90.32\% \\
\bottomrule
\end{tabular}
\caption{Case-level evaluation results.}
\label{tab:performance}
\end{table}

The perfect precision score (100\%) across both datasets demonstrates the robustness of our approach, with no false alarms triggered during testing. The system successfully identified 29 out of 36 distinct attack scenarios in the SWaT dataset and 13 out of 15 attack scenarios in the WADI dataset, achieving high recall rates of 80.56\% and 86.67\% respectively. The F1 scores of 89.25\% for SWaT and 92.86\% for WADI reflect the balanced performance between precision and recall in both environments.

Our analysis of the detection results revealed several interesting patterns. We found that certain invariants were particularly effective at detecting multiple attack scenarios. For example, as shown in Table \ref{tab:comprehensive}, $\mathbb{INV}2$ and $\mathbb{INV}6$ each detected three or more different attack types in the SWaT dataset, indicating their importance in the overall detection framework. The most successful invariants tended to be those with higher PCMCI scores, confirming our validation approach's effectiveness.
We also observed complementary detection capabilities across different invariant types. While some attacks were detected by multiple invariants, providing redundant protection, others were uniquely identified by specific invariants, as evident from the attack detection patterns in Table \ref{tab:comprehensive}. This highlights the value of maintaining diverse invariant types that cover different aspects of the system's operation.
The consistent performance across two distinct CPS environments (water treatment and water distribution) demonstrates the generalizability of our approach. By combining LLM-extracted invariants with rigorous statistical validation, we achieved robust anomaly detection capabilities that maintained high precision while delivering strong recall performance.

\subsection{Comparison with existing Solutions}

To evaluate the effectiveness of \shortname, we compared our approach with anomaly detection methods for CPS. Our comparative analysis includes methods that have been evaluated on the SWaT and WADI datasets. Table~\ref{tab:comparison} presents the performance comparison of \shortname against some existing methods. Our approach achieved perfect precision (100\%) on both datasets, completely eliminating false alarms, which are critical concerns in operational environments. While methods like STGAT-MAD~\cite{zhan2022stgat} demonstrated higher recall rates, \shortname maintained an excellent balance between precision and recall, resulting in competitive F1 scores of 89.25\% on SWaT and 92.86\% on WADI.

\begin{table*}[htbp]
  \centering
  \small
  \setlength{\tabcolsep}{20pt}
  \resizebox{0.9\textwidth}{!}{
  \begin{tabular}{l|c c c|c c c}
  \toprule
  \rowcolor[HTML]{FFFFFF}
  \textbf{Method} &
  \multicolumn{3}{c|}{\cellcolor[HTML]{FFFFFF}\textbf{SWaT}} &
  \multicolumn{3}{c}{\cellcolor[HTML]{FFFFFF}\textbf{WADI}} \\
  \rowcolor[HTML]{FFFFFF}
   &
  \textbf{Precision} & \textbf{Recall} & \textbf{F1} &
  \textbf{Precision} & \textbf{Recall} & \textbf{F1} \\ \hline
  GDN~\cite{GDN} & 99.35 & 68.12 & 81.00 & 97.50 & 40.19 & 57.00 \\ 

  NSIBF~\cite{b26} & 98.20 & 86.30 & 91.90 & 91.50 & 88.70 & 90.10 \\ 

  STGAT-MAD~\cite{zhan2022stgat} & 84.10 & 96.50 & 90.00 & 91.00 & 79.70 & 84.90 \\ 
  MAD-GAN~\cite{li2019mad} & 98.97 & 63.74 & 77.00 & 41.44 & 33.92 & 37.00 \\ 
  USAD~\cite{audibert2020usad} & 98.70 & 74.02 & 84.60 & 64.51 & 32.20 & 42.96 \\ 
  Data-driven invariant rules-I~\cite{b29} &  & 70.87 &  &  & 41.14 &  \\ 
  Data-driven invariant rules-II~\cite{b29} &  & 78.81 &  &  & 47.44 &  \\ \hline
  \textbf{\shortname} & \textbf{100.00} & 80.56 & 89.25 & \textbf{100.00} & 86.67 & 92.86 \\ \bottomrule
  \end{tabular}
  }
  \caption{Performance comparison of \shortname with other anomaly detection methods}
  \label{tab:comparison}
\end{table*}

The comparative analysis reveals some interesting patterns across the two datasets. On SWaT, most methods achieved high precision, with several approaches exceeding 95\%. The WADI dataset, however, proved more challenging for most methods, with significant performance degradation observed for approaches like MAD-GAN~\cite{li2019mad} and USAD~\cite{audibert2020usad}, whose F1 scores dropped by 40 and 41.64 percentage points, respectively. In contrast, \shortname demonstrated exceptional resilience by not only maintaining but actually improving its F1 score by 3.61 percentage points on the more complex WADI dataset. This consistency across datasets of varying complexity showcases the robust nature of our physics-based invariant approach.

Beyond competitive detection accuracy, \shortname offers practical advantages over deep learning based approaches such as GDN \cite{GDN} and MAD-GAN \cite{li2019mad}.
First, it is lightweight. LLM inference is performed only once during setup, and the detection phase requires only regressions. Because no deep neural computations run at detection time, the system remains resource-friendly and is suitable for edge devices with constrained resources. Second, it is explainable. Every alarm is tied to a specific violated invariant, giving operators clear semantic clues for diagnosis. When an incident triggers several invariants at once, the common variables reveal the likely root cause with little manual effort.
Third, it is easy to adapt to system changes. If the target CPS deployment has changes such as adding, removing, or relocating a device, we can simply delete the affected invariants and rerun the invariant extraction specifically for the modified subsystem. No costly retraining or hyper-parameter tuning is required, which keeps maintenance overhead low while preserving transparency.


\section{Related Works}\label{sec:related}

\subsection{Anomaly Detection using Physical Invariants}

Anomaly detection in CPS frequently employs physical invariants—conditions derived from underlying physical laws—to identify anomalies indicative of faults or cyber-attacks. Unsupervised machine learning has been applied to infer physical invariants in electrical substations without explicitly encoding traditional physical laws, such as Kirchhoff's laws, demonstrating flexibility in anomaly detection \cite{b1}. Another method leveraged the Local Outlier Factor (LOF) algorithm on log data to identify failures within an aquarium management system \cite{b2}. In smart grids, integrated approaches combining cyber-physical sensors with graphical models effectively distinguish between cyber and physical anomalies and subtle disturbances \cite{b3,b4,b5}. Additionally, methods that incorporate physical rules into machine learning frameworks have been effective in detecting anomalies within smart water infrastructures and distributed CPS \cite{b7,b8}. Deep learning approaches, often combined with physical invariants, significantly improved anomaly detection accuracy in advanced driver assistance systems and complex industrial processes \cite{b9,b11,b12,b13}. Privacy-preserving frameworks have also utilized physical invariants to secure sensitive data while enhancing anomaly detection performance \cite{b14,b26,b19,b27}.

\subsection{Causal Discovery for Anomaly Detection}

Causal discovery contributes to anomaly detection by revealing underlying cause-effect relationships in temporal data, improving both interpretability and robustness compared to purely statistical methods. Graph-based approaches such as MST-GAT and Graph Neural Network-based Anomaly Detection (GDN) utilize attention mechanisms to dynamically capture causal dependencies, significantly enhancing anomaly attribution \cite{MST-GAT,GDN}. Granger causality-based methods, including deep learning enhancements, effectively model complex interactions among sensors, particularly in high-dimensional data scenarios \cite{GrangerDeep,FuSAGNet}. Structural causal models (SCMs) further aid in distinguishing genuine causal relationships from mere correlations, crucial for accurate anomaly diagnosis within CPS \cite{causal_disc}. The integration of causal discovery with deep anomaly detection continues to enhance detection accuracy, interpretability, and decision-making efficacy.

\subsection{LLM-based Anomaly Detection}

Recent advancements have integrated LLMs into anomaly detection, demonstrating notable successes in video surveillance (e.g., VAD-LLaMA) and various CPS applications \cite{b33,b31}. In CPS contexts, LLMs facilitate automatic extraction of physical invariants, significantly streamlining monitoring processes and enhancing scalability in industrial control systems and smart grids by combining real-time sensor inputs with historical data analysis \cite{b32,alam2024assess,b30,rashvand2024enhancing}. Models such as GPT-4 further bolster CPS resilience by addressing context grounding and formal verification challenges, which are essential for deployment in deterministic environments \cite{b15,b16,b18}. Nonetheless, deploying LLMs in anomaly detection demands robust assurance frameworks to tackle issues such as data dependency, limited generalizability, and computational complexity, ensuring reliable integration and operation within CPS environments \cite{b19,b20,zibaeirad2024vulnllmeval}.


\section{Conclusion}\label{sec:conclusion}
In this work, we explore the potential of LLMs for anomaly detection in CPS. We introduce a novel method that automates the extraction of physical invariants by leveraging semantic information from CPS documentation, a resource often overlooked in existing solutions. Our method effectively augments traditional data-driven approaches through the integration of semantic insights and empirical data analysis. Additionally, we develop an efficient extraction workflow and an innovative validation algorithm designed to improve scalability and accuracy in CPS anomaly detection. Our comprehensive evaluation on two public datasets shows excellent accuracy, particularly in minimizing false alarms. Our method can accurately raise alarms upon the occurrence of attack cases, demonstrating the feasibility of this innovative approach.

\section{Ethics Considerations}
This work does not involve any personally identifiable information (PII), real-world physical systems, or human participants. All experiments are conducted on publicly accessible datasets. Therefore, no ethical concerns are associated with this study.

\bibliographystyle{IEEEtran}
\bibliography{NDSS}

\clearpage
\section*{Appendix}
\subsection*{A. Full List of SWaT and WADI Invariants}

\begin{center}
\centering
\small
\begin{tabular}{c| |c|p{15cm}}
\toprule
\textbf{Dataset} & \textbf{ID} & \textbf{Invariant} \\
\hline

\multirow{38}{*}{\rotatebox[origin=c]{90}{SWaT}} & $\mathbb{INV}1$ & $\frac{d}{dt}\text{LIT101} = C_1 \cdot \text{FIT101} + C_2 \cdot \text{FIT201}$ \\
\cline{2-3}
& $\mathbb{INV}2$ & $\frac{d}{dt}\text{LIT401} = C_3 \cdot \text{FIT301} + C_4 \cdot \text{FIT401}$ \\
\cline{2-3}
& $\mathbb{INV}3$ & $\frac{d}{dt}\text{LIT401} = C_5 \cdot \text{FIT301}$ \\
\cline{2-3}
& $\mathbb{INV}4$ & $\text{FIT501} = C_6 \cdot \text{FIT502} + C_7 \cdot \text{FIT503}$ \\
\cline{2-3}
& $\mathbb{INV}5$ & $\text{FIT501} = C_8 \cdot \text{FIT502}$ \\
\cline{2-3}
& $\mathbb{INV}6$ & $\text{FIT501} = C_9 \cdot \text{FIT503}$ \\
\cline{2-3}
& $\mathbb{INV}7$ & $\text{AIT201} \cdot \text{FIT201} = C_{10} \cdot \text{P201}$ \\
\cline{2-3}
& $\mathbb{INV}8$ & $\text{FIT101} = C_{11} \cdot \text{MV101}$ \\
\cline{2-3}
& $\mathbb{INV}9$ & $\text{DPIT301} = C_{12} \cdot \text{FIT301}$ \\
\cline{2-3}
& $\mathbb{INV}10$ & $\text{FIT201} = C_{13} \cdot \text{MV201}$ \\
\cline{2-3}
& $\mathbb{INV}11$ & $\text{AIT402} = C_{14} \cdot \text{FIT401}$ \\
\cline{2-3}
& $\mathbb{INV}12$ & $\frac{d}{dt}\text{LIT301} = C_{15} \cdot \text{FIT201} + C_{16} \cdot \text{P301} + C_{17} \cdot \text{P302}$ \\
\cline{2-3}
& $\mathbb{INV}13$ & $\text{AIT501} = C_{18} \cdot \text{FIT501}$ \\
\cline{2-3}
& $\mathbb{INV}14$ & $\text{DPIT301} = C_{19} \cdot \text{FIT301}^2$ \\
\cline{2-3}
& $\mathbb{INV}15$ & $\frac{d}{dt}\text{LIT101} = C_{20} \cdot \text{FIT101} + C_{21} \cdot \text{P101} + C_{22} \cdot \text{P102}$ \\
\cline{2-3}
& $\mathbb{INV}16$ & $\frac{d}{dt}\text{LIT101} = C_{23} \cdot \text{P101}$ \\
\cline{2-3}
& $\mathbb{INV}17$ & $\text{AIT203} = C_{24} \cdot \text{P205} + C_{25} \cdot \text{P206}$ \\
\cline{2-3}
& $\mathbb{INV}18$ & $\text{AIT203} = C_{26} \cdot \text{P205}$ \\
\cline{2-3}
& $\mathbb{INV}19$ & $\text{AIT203} \cdot \text{FIT201} = C_{27} \cdot \text{P205} + C_{28} \cdot \text{P206}$ \\
\cline{2-3}
& $\mathbb{INV}20$ & $\frac{d}{dt}\text{LIT101} = C_{29} \cdot \text{MV101} + C_{30} \cdot \text{FIT101}$ \\
\cline{2-3}
& $\mathbb{INV}21$ & $\frac{d}{dt}\text{LIT301} = C_{31} \cdot \text{FIT201} + C_{32} \cdot \text{FIT301}$ \\
\cline{2-3}
& $\mathbb{INV}22$ & $\frac{d}{dt}\text{LIT101} = C_{33} \cdot \text{FIT101} + C_{34} \cdot \text{P101}$ \\
\cline{2-3}
& $\mathbb{INV}23$ & $\text{MV101} \cdot \text{FIT101} = C_{35} \cdot \text{P101}$ \\
\cline{2-3}
& $\mathbb{INV}24$ & $\text{FIT101} = C_{36} \cdot \text{LIT101} + C_{37} \cdot \text{MV101}$ \\
\cline{2-3}
& $\mathbb{INV}25$ & $\text{FIT501} = C_{38} \cdot \text{FIT502} + C_{39} \cdot \text{FIT503} + C_{40} \cdot \text{FIT504}$ \\
\cline{2-3}
& $\mathbb{INV}26$ & $\text{DPIT301} = C_{41} \cdot \text{FIT301} + C_{42} \cdot \text{LIT301}$ \\
\cline{2-3}
& $\mathbb{INV}27$ & $\text{FIT401} = C_{43} \cdot \text{FIT501} + C_{44} \cdot \text{FIT502}$ \\
\cline{2-3}
& $\mathbb{INV}28$ & $\text{FIT401} = C_{45} \cdot \text{FIT502}$ \\
\cline{2-3}
& $\mathbb{INV}29$ & $\text{FIT401} = C_{46} \cdot \text{FIT501}$ \\
\cline{2-3}
& $\mathbb{INV}30$ & $\text{PIT501} = C_{47} \cdot \text{FIT501} + C_{48} \cdot \text{FIT502}$ \\
\cline{2-3}
& $\mathbb{INV}31$ & $\frac{d}{dt}\text{LIT401} = C_{49} \cdot \text{FIT401} + C_{50} \cdot \text{P401} + C_{51} \cdot \text{P402}$ \\
\cline{2-3}
& $\mathbb{INV}32$ & $\text{P203} = C_{52} \cdot \text{P205} + C_{53} \cdot \text{AIT202} + C_{54} \cdot \text{AIT203}$ \\
\cline{2-3}
& $\mathbb{INV}33$ & $\text{PIT501} = C_{55} \cdot \text{FIT503} + C_{56} \cdot \text{FIT504}$ \\
\cline{2-3}
& $\mathbb{INV}34$ & $\text{AIT504} = C_{57} \cdot \text{AIT503}$ \\
\cline{2-3}
& $\mathbb{INV}35$ & $\text{AIT402} = C_{58} \cdot \text{UV401} + C_{59} \cdot \text{FIT401}$ \\
\cline{2-3}
& $\mathbb{INV}36$ & $\text{FIT601} = C_{60} \cdot \text{FIT503}$ \\
\cline{2-3}
& $\mathbb{INV}37$ & $\frac{d}{dt}\text{LIT101} = C_{61} \cdot \text{MV101} \cdot \text{FIT101} + C_{62} \cdot \text{P101} + C_{63} \cdot \text{P102}$ \\
\cline{2-3}
& $\mathbb{INV}38$ & $\text{DPIT301} = C_{64} \cdot \text{P301} + C_{65} \cdot \text{FIT301}^2$ \\
\hline
\multirow{12}{*}{\rotatebox[origin=c]{90}{WADI}} & $\mathbb{INV}39$ & $\text{2\_PIT\_003\_PV} = C_{66} \cdot \text{2\_FIT\_003\_PV}$ \\
\cline{2-3}
& $\mathbb{INV}40$ & $\text{2\_MCV\_007\_CO} = C_{67} \cdot \text{2\_MCV\_101\_CO} + C_{68} \cdot \text{2\_MCV\_201\_CO} + C_{69} \cdot \text{2\_MCV\_301\_CO} + C_{70} \cdot \text{2\_MCV\_401\_CO} + C_{71} \cdot \text{2\_MCV\_501\_CO} + C_{72} \cdot \text{2\_MCV\_601\_CO} + C_{73} \cdot \text{2\_FQ\_101\_PV} + C_{74} \cdot \text{2\_FQ\_201\_PV} + C_{75} \cdot \text{2\_FQ\_301\_PV} + C_{76} \cdot \text{2\_FQ\_401\_PV} + C_{77} \cdot \text{2\_FQ\_501\_PV} + C_{78} \cdot \text{2\_FQ\_601\_PV}$ \\
\cline{2-3}
& $\mathbb{INV}41$ & $\text{2\_PIT\_001\_PV} = C_{79} \cdot \text{2\_LT\_002\_PV}$ \\
\cline{2-3}
& $\mathbb{INV}42$ & $\text{2\_DPIT\_001\_PV} = C_{80} \cdot \text{2\_PIT\_001\_PV} + C_{81} \cdot \text{2\_PIT\_002\_PV}$ \\
\cline{2-3}
& $\mathbb{INV}43$ & $\text{2\_PIT\_003\_PV} = C_{82} \cdot \text{2\_P\_003\_STATUS}$ \\
\cline{2-3}
& $\mathbb{INV}44$ & $\text{2\_PIT\_001\_PV} = C_{83} \cdot \text{2\_PIT\_002\_PV} + C_{84} \cdot \text{2\_FIT\_002\_PV} \cdot \text{2\_FIT\_002\_PV}$ \\
\cline{2-3}
& $\mathbb{INV}45$ & $\text{2\_PIT\_001\_PV} = C_{85} \cdot \text{2\_PIT\_002\_PV} + C_{86} \cdot \text{2\_FIT\_002\_PV} \cdot \text{2\_FIT\_002\_PV} \cdot \text{2\_MCV\_007\_CO}$ \\
\cline{2-3}
& $\mathbb{INV}46$ & $\text{2\_FIT\_002\_PV} = C_{87} \cdot \text{2\_FIT\_003\_PV} + C_{88} \cdot \text{2\_FIC\_101\_PV} + C_{89} \cdot \text{2\_FIC\_201\_PV} + C_{90} \cdot \text{2\_FIC\_301\_PV} + C_{91} \cdot \text{2\_FIC\_401\_PV} + C_{92} \cdot \text{2\_FIC\_501\_PV} + C_{93} \cdot \text{2\_FIC\_601\_PV}$ \\
\cline{2-3}
& $\mathbb{INV}47$ & $\text{1\_FIT\_001\_PV} = C_{94} \cdot \text{1\_MV\_001\_STATUS}$ \\
\cline{2-3}
& $\mathbb{INV}48$ & $\text{2\_PIT\_001\_PV} = C_{95} \cdot \text{2\_LT\_001\_PV} + C_{96} \cdot \text{2\_LT\_002\_PV}$ \\
\cline{2-3}
& $\mathbb{INV}49$ & $\frac{d}{dt}\text{2\_FQ\_401\_PV} = C_{97} \cdot \text{2\_FIC\_401\_PV}$ \\
\cline{2-3}
& $\mathbb{INV}50$ & $\text{1\_AIT\_001\_PV} = C_{98} \cdot \text{2A\_AIT\_001\_PV} + C_{99} \cdot \text{3\_AIT\_001\_PV}$ \\
\bottomrule
\end{tabular}
\end{center}

\end{document}